\documentstyle[prl,aps,preprint]{revtex}
\tightenlines

\newcommand{\be}{\begin{equation}}
\newcommand{\ee}{\end{equation}}
\newcommand{\bea}{\begin{eqnarray}}
\newcommand{\eea}{\end{eqnarray}}
\def\fct#1{\mathop{\rm #1}}	
\def\im{\fct{Im}}
\newcommand{\Eq}{Eq. \ref}

\newcommand{\Cz}{\mbox{\boldmath $C$}}
\def\eps{\varepsilon}
\def\<{\langle} 				
\def\>{\rangle} 				

\begin{document}
\preprint{Submitted to {\em Physical Review Letters}}


\title{Pseudo-time Schr\"odinger 
equation with absorbing potential for quantum scattering calculations.}

\author{Arnold Neumaier$^1$ and Vladimir A. Mandelshtam$^2$}

\address{$^1$ Institut f\"ur Mathematik, Universit\"at Wien
Strudlhofgasse 4, A-1090 Wien, Austria;\\ email: neum@cma.univie.ac.at; 
WWW: http://solon.cma.univie.ac.at/$\sim$neum/}

\address{$^2$ Chemistry Department, 
University of California at Irvine, Irvine, CA 92697, USA; 
email: mandelsh@uci.edu}

\date{\today}

\maketitle

\abstract{  
The Schr\"odinger equation $(H\psi)(r)=(E+u_EW(r))\psi(r)$ with an
energy-dependent complex absorbing potential $-u_E W(r)$, associated 
with a scattering system, can be reduced for a special choice of $u_E$ 
to a harmonic inversion problem of a discrete pseudo-time correlation 
function $y(t)=\phi^{\rm T} U^t\phi$. An efficient formula for
Green's function matrix elements is also derived. Since the exact 
propagation up to time $2t$ can be done with only $\sim t$ real 
matrix-vector products, this gives an unprecedently efficient scheme
for accurate calculations of quantum spectra for possibly very
large systems. 
}} 



\newpage

\noindent 
{\bf Complex absorbing potentials.}
The spectrum of a quantum scattering system can be characterized by
solving the boundary value problem associated with the Schr\"odinger equation,
\be\label{ab.1}
   (H\psi)(r)=E\psi(r).
\ee
The {\it bound states} then have real energies $E$, with solutions
$\psi(r)$ exponentially localized in space.
The {\it resonance states} (Siegert states \cite{siegert}) have complex energies with $\im E\leq 0$. They behave like 
bound states in some compact subset $\Omega$ of the configuration
space, but eventually grow exponentially
outside of $\Omega$, due to the outgoing asymptotic boundary conditions.

By introduction of a so-called {\it optical (or absorbing)
potential} $-u W(r)$ with $\im u>0$ and real $W(r)\geq 0$, that
vanishes for $r\in\Omega$ and smoothly grows outside $\Omega$, 
the solutions $\psi(r)$ of 
\be\label{ab.2}
   (H\psi)(r)=(E+u W(r))\psi(r)
\ee
are damped outside $\Omega$, the physically relevant region \cite{Jolicard}. 
This forces them to behave like bound states everywhere without
significantly affecting the energies $E$. In this framework
the physically relevant part of the system is, therefore, 
dissipative and satisfies (\ref{ab.1}) only for $r\in\Omega$.
Moreover, a general multichannel scattering problem 
can be considered with a numerically 
convenient form of $W(r)$, independent of
the choice of coordinate system.

Although to satisfy  $\im E\leq 0$ one only needs
$\im u\ge 0$, traditionally one simply puts $u=i$ and gets 
the nonhermitian eigenvalue problem $(H-iW)\psi=E\psi$. The latter is
generally much easier to handle numerically 
than the boundary value problem (\ref{ab.1}). As we shall see, 
energy-dependent choices $u=u_E$ are particularly useful. 

The introduction of the absorbing potential leads to the {\it damped
Green's function} \cite{NeuhBaer,SeidMill,RisM}
\be\label{ab.GW}
G_W(E):=(E-H+u_E W)^{-1}.
\ee
Under suitable conditions on $u_EW(r)$, it is probably 
possible to prove, similar as in ref.~\cite{RisM} for 
the traditional case $u_E=i$, that $G_W(E)$ converges for any real
$E$ (and also for $\im E\ge 0$) 
weakly to the ordinary Green's function
\[
G(E)=\lim_{\eps\downarrow 0} (E-H+i\eps)^{-1}.
\]
Practically, one usually needs to evaluate only certain matrix elements 
$\phi^{\rm T}G(E)\psi$, the basic numerical 
objects of quantum physics 
from which most other quantities of interest (scattering amplitudes, 
reaction rates, etc.) can be computed (see, e.g.,
refs.~\cite{NeuhBaer,SeidMill,TanWeeks,Kourivar}). 
If both $\phi$ and $\psi$ have support in $\Omega$, they are well
approximated by $\phi^{\rm T}G_W(E)\psi$.

Unfortunately, for very large systems with high density of states one 
may encounter numerical difficulties when trying to diagonalize a large
nonhermitian matrix $H-u_E W$ or solve the linear system 
$(E-H+u_E W)X(E)=\psi$ at many values of $E$ using
general iterative techniques for nonhermitian matrices.
However, one can devise alternative iterative
techniques to solve (\ref{ab.2}) by exploiting the special
structure of the quantum scattering problem.

\bigskip
  From now on, we assume that the Hilbert space is discretized so 
that the states are vectors $\psi\in\Cz^K$ and $H$, $W$ are
real symmetric $K\times K$ matrices, $W$ 
diagonal, as, e.g., in the case of a discrete variable 
representation \cite{DVR}.

To simplify the following equations we further assume 
without loss of generality that the discretized Hamiltonian is shifted 
and scaled so that $|\<H\>_{\psi}|\leq 1$ for any state
$\psi$, where we have defined the expectation value as
$\<H\>_{\psi}:=\psi^*H\psi/ \psi^*\psi$ and $^*$ denotes conjugate
transposition. Such scaling is implemented routinely in the framework
of the Chebyshev polynomial expansion.

We now consider the special choice
\be\label{ab.3}
   u_E=E+ i\sqrt{1-E^2}, \ \ \mbox{or}\ \ E(u)={1+u^2\over 2u},
\ee 
which after insertion into (\ref{ab.2}) gives a nonlinear 
eigenvalue problem for $u=u_E$, more useful than (\ref{ab.2}):
\be\label{ab.4}
   H\psi={1+u^2 D\over 2u}\psi\ \ \mbox{with}\ \ D=1+2W.
\ee
We may think of this equation as an eigenvalue problem 
\be\label{ab.4a}
H\psi=E(u)\psi
\ee
involving an operator-valued $u$-dependent energy 
\be\label{ab.4c}
E(u)={1+u^2 D\over 2u}
\ee
that reduces in $\Omega$ (where $D(r)=1$)
to the constant (\ref{ab.3}). 

In ref.~\cite{ManT2} a similar nonlinear eigenvalue problem with 
$E(u)=(D^{-1}+u^2D)/2u$
was implicitly encountered leading to a numerical scheme to compute, 
e.g., the complex resonance energies by {\it harmonic inversion} of a 
``discrete-time'' correlation
function, generated by a damped Chebyshev recursion.
Here, related results are derived rigorously and in a more
general framework. In particular,
a {\it pseudo-time Schr\"odinger equation} is derived
that allows one to achieve a substantial numerical saving compared
to the previous works.

The eigenpairs $(u_k,\psi_k)$ of (\ref{ab.4}) can be used to evaluate
the physically interesting quantities (e.g., the complex resonance 
energies $E_k$, scattering amplitudes, etc.). However,
because of the nonlinearity they have somewhat different properties 
from those of the regular nonhermitian eigenvalue problem, which we 
now proceed to derive.

\bigskip 
\noindent 
{\bf Theorem 1: Completeness and Orthogonality.}
The nonlinear eigenvalue problem (\ref{ab.4}) has at most $2K$ distinct
eigenvalues $u_k$. Thus, there are at most $K$ physical
eigenvalues $E_k$ of (\ref{ab.2}) with $\im E_k\geq 0$. \\
If there are $2K$ distinct eigenvalues $u_1,\dots,u_{2K}$
with associated eigenvectors $\psi_k$ satisfying (\ref{ab.4}),
then for any vector $\phi_{\alpha}\in\Cz^K$ there is a set of $2K$ numbers
$\theta_{\alpha k}$ satisfying the {\it completeness relations}
\be\label{ab.expan}
   \phi_{\alpha}=\sum_{k=1}^{2K}\theta_{\alpha k}\psi_k,\ \ 
   0=\sum_{k=1}^{2K}\theta_{\alpha k}u_k\psi_k.
\ee
Furthermore, $\psi_k$ satisfy the {\it orthogonality 
relations}
\be\label{ab.orth}
\psi_j^{\rm T}(1-u_ju_kD)\psi_k = \delta_{jk}
\ee
for all $j\ne k$. If the eigenvectors can be normalized such that 
(\ref{ab.orth}) also holds for $j=k$, then (\ref{ab.expan}) holds with
\be\label{ab.coef}
\theta_{\alpha k}=\psi_k^{\rm T}\phi_{\alpha}
\ee
\noindent 
{\it Proof.}
Let
\[
U={0 \ \ \ \ \ \ \ \ \ \ I \choose
           -D^{-1}  \ \ D^{-1}(2H)}.
\]
where $I$ denotes the $K\times K$ unit matrix.
The ordinary eigenvalue problem
\be\label{ab.9}
   U\hat\psi=u\hat\psi
\ee
yields, with $\hat\psi={\psi\choose\psi'}$,  
\[
\psi'=u\psi,~~~D^{-1} \psi+u^2\psi-uD^{-1}(2H)\psi=0.
\]
After multiplication by $D/2u$, we find (\ref{ab.4}). 
In particular, the eigenpairs $(u_k,\hat\psi_k)$ of $U$ satisfy
\be\label{ab.11}
   \hat\psi_k={\psi_k\choose u_k\psi_k},
\ee
where $(u_k,\psi_k)$ is an eigenpair of (\ref{ab.4}). Since conversely, 
any such eigenpair determines an eigenpair of $U$,
the nonlinear eigenvalue problem (\ref{ab.4}) has at most $2K$ distinct
eigenvalues. If there are $2K$ distinct eigenvalues $u_1,\dots,u_{2K}$,
the matrix $U$ is diagonalizable, and there is a basis 
$\hat\psi_1,\dots,\hat\psi_{2K}$ of eigenvectors of the form 
(\ref{ab.11}). Therefore, we may write 
\[
   {\phi_{\alpha} \choose 0}=\sum_{k=1}^{2K}\theta_{\alpha k}\hat\psi_k=
   {\sum\theta_{\alpha k}\psi_k\choose\sum\theta_{\alpha k}u_k\psi_k}
\]
with uniquely determined coefficients $\theta_{\alpha k}$. This gives 
(\ref{ab.expan}).

Using (\ref{ab.4}) and the symmetry of $H$ and $D$ we may compute 
$\psi_j^{\rm T}H\psi_k$ in two different ways:
\[
\psi_j^{\rm T}H\psi_k
=\psi_j^{\rm T}{1+u_k^2D\over 2u_k} \psi_k
=\psi_k^{\rm T}{1+u_j^2D\over 2u_j} \psi_j.
\]
For $j\ne k$, by assumption $u_j\neq u_k$. Thus, we may take 
difference, multiply by $2u_ju_k/(u_k-u_j)$, and find 
(\ref{ab.orth}). 
For $j=k$, we may achieve (\ref{ab.orth}) by
normalizing the eigenvectors, provided that the left hand side of
(\ref{ab.orth}) does not vanish. Using (\ref{ab.orth}) and 
(\ref{ab.expan}) 
we find (\ref{ab.coef}):
\bea\nonumber
\theta_{\alpha k}&=& {\sum}_j \theta_{\alpha j}\delta_{jk}
={\sum}_j \theta_{\alpha j}\psi_j^{\rm T}(1-u_ju_kD)\psi_k\\\nonumber
&=& \left(\sum \theta_{\alpha j}\psi_j\right)^{\rm T}\psi_k
-u_k\left(\sum \theta_{\alpha j}u_j\psi_j\right)^{\rm T}D\psi_k 
\\\nonumber &=& \phi_{\alpha}^{\rm T}\psi_k=\psi_k^{\rm T}\phi_{\alpha}.
\eea

\bigskip
We now consider an eigenpair $(u,\psi)$ of (\ref{ab.4}) with 
$\psi^*\psi\ne 0$. 
Multiplying (\ref{ab.4}) by $2u\psi^*$ gives the quadratic equation
\be\label{ab.5}
   u^2\<D\>_{\psi}-2u\<H\>_{\psi}+1=0.
\ee
The solutions of (\ref{ab.5}) are
\be\label{ab.6}
   u={\<H\>_{\psi}\pm i\sqrt{\<D\>_{\psi}-\<H\>_{\psi}^2}\over \<D\>_{\psi}}.
\ee
Since $\<D\>=1+2\<W\>_{\psi}\ge 1$ and  $|\<H\>_{\psi}|\leq 1$, the square root
is real and
\be\label{ab.u2}
|u|^2={\<H\>_{\psi}^2+\<D\>_{\psi}-\<H\>_{\psi}^2\over \<D\>_{\psi}^2}
     ={1\over \<D\>_{\psi}}\leq 1.
\ee
Thus, $u$ is a complex number lying in the unit disk. Moreover, 
$|u|=1$ iff $\<W\>_{\psi}=0$, i.e., iff $\psi$ has support in 
$\Omega$, which is the case for the bound states. The states with
$|u|\sim 1$ correspond to the narrow resonances.
Due to (\ref{ab.6}) the solutions of (\ref{ab.4}) come in complex 
conjugate pairs $(u,\psi)$ and $(\bar u,\bar \psi)$. The 
physically relevant eigenenergies with $\im E\leq 0$ come from  
$u$ with $\im u\geq 0$.

\bigskip
Note that a similar analysis of a quadratic eigenvalue problem was 
carried out in ref.~\cite{TPRL}, arising from the use of 
the Bloch operator $L$, rather than an absorbing potential. There,
equation (\ref{ab.4a}) is considered with $E(u)=(iuL+u^2 I)/2$,
and $u$ is a momentum variable close to the real axis, instead of 
a number close to the unit circle. However, 
this equation can only be used for less general, single-channel
scattering problems and, besides, it is hard to solve efficiently
using iterative techniques.

\bigskip 
\noindent 
{\bf Reduction to a harmonic inversion problem.}
Consider the {\it pseudo-time Schr\"odinger equation} defined
by the recurrence
\be\label{ab.12}
   \phi(t)=D^{-1}(2H\phi(t-1)-\phi(t-2))\ \ (t=2,3,...)
\ee
with $\phi(0)=\phi_0$ and  $\phi(1)=0$.
(This choice of initial conditions is most convenient, although other
initial conditions with $\phi(1)\ne 0$ yield analogous results. 
A similar 3-term-recurrence with another choice of special 
initial conditions leading to ``modified Chebyshev recurrence'' was
considered in refs.~\cite{ManT2,ManT}.) 
Since $D$ is diagonal and matrix-vector products $H\phi$ are usually 
cheap to form, $\phi(0),\dots,\phi(T)$ are computable using $O(KT)$ 
operations and a few vectors stored at a time. If the initial vector 
$\phi_0$ is real, only real arithmetic is needed. 

By Theorem 1, we can write (\ref{ab.12}) as 
\[
{\phi(t)\choose \phi(t+1)}=U^t{\phi_0\choose 0}=\sum_{k=1}^{2K}\theta_{0k}
u_k^t{\psi_k\choose u_k\psi_k}
\]
and, therefore,
\be\label{ab.13}
   \phi(t)=\sum_{k=1}^{2K}\theta_{0k}u_k^t\psi_k.
\ee
This {\it power expansion} is very important, and is analogous to
the (physical time) expansion
$\phi(t)=e^{-itH}\phi(0)=\sum_{k=1}^{K}\theta_ke^{-itE_k}\psi_k$
for the solutions of the standard time-dependent Schr\"odinger equation.
It allows one to reap all the benefits of time-dependent 
methods (see, e.g., refs.~\cite{Heller,TD_approaches,TanWeeks}) 
without having to deal with the time-dependent 
Schr\"odinger equation, which is hard to solve
accurately at long times $t$ in the case of nonhermitian Hamiltonian. 
Instead, only the much more benign and numerically very 
stable equation (\ref{ab.12}) must be solved.

By (\ref{ab.13}), the pseudo-time cross-correlation function
\be\label{ab.14}
   y_{\alpha}(t):=\phi_{\alpha}^{\rm T}\phi(t)\ \ (t=0,1\dots)
\ee
of a state $\phi_{\alpha}$ has the form
\be\label{ab.15}
   y_{\alpha}(t)=\sum_{k=1}^{2K}d_{\alpha k} u_k^t
\ee 
with 
\be\label{ab.dk}
d_{\alpha k}=
\phi_{0}^{\rm T}\psi_k\psi_k^{\rm T}\phi_{\alpha}
=\theta_{0k}\theta_{\alpha k}. 
\ee
This reduces the nonlinear 
eigenvalue problem (\ref{ab.4}) to solving the {\it harmonic inversion 
problem}, i.e., to finding the spectral parameters
$(u_k,d_{\alpha k})$ ($k=1,\dots,2K$) satisfying (\ref{ab.15}) for the 
sequence $y_{\alpha}(t)$ computed by (\ref{ab.12}) and (\ref{ab.14}). 
Since by (\ref{ab.u2}) the sequence $y_{\alpha}(t)$ is bounded and
the spectral mapping (\ref{ab.3}) moves the
physically relevant eigenvalues $u_k$ close to the unit circle,
this is an efficiently tractable problem, even in very large 
dimensions \cite{Neuh95,ManT2}.

\bigskip
\noindent
{\bf Time doubling of an autocorrelation function.}
As is well known, a true time autocorrelation function at time $t$ 
can be computed by solving the time-dependent Schr\"odinger
equation up to time $t/2$, since one can use
\[
C(t):=\phi^{\rm T}e^{-iHt}\phi=(e^{-iHt/2}\phi)^{\rm T}(e^{-iHt/2}\phi).
\]
For the Chebyshev autocorrelation function 
$c(t):=\phi_\alpha^{\rm T}\phi(t)$, based on (\ref{ab.12}) with $D=I$
and the initial conditions $\phi(0)=\phi_0$, $\phi(1)=H\phi_0$,
a factor of two saving is also well known 
(see, e.g., the discussion in ref.~\cite{ManT2}):
\[
c(2t)=2\phi(t)^{\rm T}\phi(t)-c(0),\, \ 
c(2t+1)=2\phi(t)^{\rm T}\phi(t+1)-c(1).
\]
This expression was used in \cite{LiG}
for resonance computation implementing a damped Chebyshev recursion. 
However, being approximate, it only worked for sufficiently narrow 
resonances. In the present framework, we can write the pseudo-time 
cross-correlation function as
\[
y_{\alpha}(t) = {\phi_\alpha\choose 0}^{\rm T}U^{t+s}{\phi_0\choose 0}=
\left\{ {\phi_\alpha\choose 0}^{\rm T} U^{t} \right\}
\left\{ U^{s}{\phi_0\choose 0}\right\},
\]
which suggests that an exact doubling scheme exists. This is now
derived for the autocorrelation function as only $\phi_0$ is propagated.

\bigskip
\noindent
{\bf Theorem 2: The Doubling Scheme.}\hskip 0.2in
For vectors $\phi(t)$ and $\phi(s)$ satisfying the pseudo-time
Schr\"odinger equation (\ref{ab.12}) with initial conditions 
$\phi(0)=\phi_0$, $\phi(1)=0$, the autocorrelation function
$y_0(t):=\phi_0^{\rm T}\phi(t)$ satisfies
\[
y_0(s+t)=\phi(s)^{\rm T}\phi(t)-\phi(s+1)^{\rm T}D\phi(t+1)
=:z(s,t).
\]

\bigskip
\noindent
{\it Proof: }
This follows from the power expansion (\ref{ab.13}) and the 
orthogonality relations (\ref{ab.orth}):
\bea\nonumber
   z(s,t) &=& \sum_{j,k=1}^{2K}\theta_{0j}\theta_{0k}u_j^su_k^t
   (\psi_l^{\rm T}\psi_k-u_ju_k\psi_l^{\rm T}D\psi_k)\\\nonumber
   &=& \sum_{k=1}^{2K}\theta_{0k}^2u_k^{s+t} = y_0(s+t).
\eea

\bigskip
Hardly any additional storage will be needed if the sequence 
$y_0(t)\ (t=0,...,2T-2)$ is generated by
\be\label{ab.y2t} 
y_0(2t)=z(t,t),\ \ y_0(2t-1)=z(t,t-1),
\ee
concurrently with the computation of $\phi(t)$ using
$t=0,...,T$. 
In exact arithmetic the harmonic inversion of the doubled sequence 
$y_0(t)$ will give the exact results if $T> 2K$, thus, using only 
$T\sim 2K$ of matrix-vector products. However, this is impractical as 
it would formally require to solve a $T\times T$ eigenvalue problem.
To reduce the computational burden and to maintain numerical 
stability the eigenvalues are extracted very efficiently in a small 
Fourier subspace by the {\it Filter Diagonalization Method} 
\cite{Neuh95,ManT2}. In this case, the required length $2T$ of the 
doubled sequence needed to converge an eigenenergy $E_k$ (cf. \Eq{ab.3})
will be defined by the locally averaged density of states
$\rho(E)$ for $E_k\sim E$ \cite{ManT2}.

\bigskip
\noindent
{\bf Theorem 3: The Green's function matrix elements.}
Under the assumptions of Theorem 1, let $\phi(t)$
be a solution of the pseudo-time Schr\"odinger equation (\ref{ab.12}) 
with initial conditions $\phi(0)=\phi_0,\ \phi(1)=0$.
Then the matrix elements of the
damped Green's function (\ref{ab.GW}) with
$u_E=E+i\sqrt{1-E^2}$ are
\be\label{ab.GWab}
\phi_{\beta}^{\rm T}G_W(E)\phi_{\alpha}=\sum_{k=1}^{2K} 
\frac{d_{\beta k}d_{\alpha k}}{d_{0k}}\frac{2u_Eu_k}{u_k-u_E},
\ee
where the three sets of spectral parameters $\{d_{\beta k},u_k\}$, 
$\{d_{\alpha k},u_k\}$ and $\{d_{0k},u_k\}$ (with identical 
eigenvalues $u_k$) satisfy the harmonic inversion problem 
(\ref{ab.15}) for the cross-correlation functions $y_{\beta}(t)$,
$y_{\alpha}(t)$ and $y_0(t)$, respectively.

\bigskip
\noindent
{\it Proof.} 
For an eigenpair $(u_k,\psi_k)$ of (\ref{ab.4}) we can write
\[
(E-H+u_EW)\psi_k 
={u_k-u_E\over 2 u_ku_E}(1-u_ku_ED)\psi_k.
\]
Then (\ref{ab.expan}) and (\ref{ab.coef}) imply
\bea\nonumber
&& (E-H+u_EW)\sum_k {2u_Eu_k\theta_{\alpha k}\over u_k-u_E} \psi_k
\\ \nonumber &=& \sum_k\theta_{\alpha k}\psi_k-u_ED\left(\sum_k \theta_{\alpha k} u_k
\psi_k\right) 
= \sum_k\theta_{\alpha k}\psi_k = \phi_{\alpha}.
\eea
Multiplying this by $\phi_{\beta}^{\rm T}(E-H+u_EW)^{-1}$ and using
$\theta_{\beta k}=\phi_{\beta}^{\rm T}\psi_k$ we obtain
\be\label{ab.GWtheta}
\phi_{\beta}^{\rm T}G_W(E)\phi_{\alpha}=\sum_{k=1}^{2K} 
\theta_{\beta k}\theta_{\alpha k}\frac{2u_Eu_k}{u_k-u_E},
\ee
Now replacement of
$\theta_{\beta k}\theta_{\alpha k}$ by 
$d_{\beta k}d_{\alpha k}/d_{0k}$ gives (\ref{ab.GWab}).

Note that (\ref{ab.GWtheta}) also gives an explicit expression for
the damped Green's function in terms of the eigenpairs:
\be
G_W(E)=\sum_{k=1}^{2K} 
\frac{2u_Eu_k}{u_k-u_E}\psi_k\psi_k^{\rm T},
\ee

A formula similar to (\ref{ab.GWab}) was obtained (without a rigorous 
derivation) in ref.~\cite{Man}, in the framework of the damped
Chebyshev recursion in place of (\ref{ab.12}). However, here, due to 
the doubling scheme (\ref{ab.y2t}), only half the 
number of matrix-vector products will be needed
to obtain the same amount of information.

Thus, we have a very efficient and stable method to extract the
complete spectral and dynamical information of a general
(multichannel) quantum scattering
system using a minimal number of matrix-vector products. This will be
demonstrated numerically in a forthcoming publication.

\bigskip
\noindent
{\bf Acknowledgement.}\hskip 0.2in
V.A.M. acknowledges the NSF support, grant CHE-9807229.



\end{document}